\documentclass[final,3p,times]{elsarticle}

\usepackage{lineno,hyperref}
\usepackage{lineno,hyperref}
\usepackage{graphicx}
\usepackage{dcolumn}
\usepackage{bm}
\usepackage{float}
\usepackage{epsfig}
\usepackage{booktabs}
\usepackage{subfigure}
\usepackage{graphics}
\usepackage{amssymb}
\usepackage{amsmath}
\usepackage{array}
\usepackage{color}
\usepackage{booktabs}
\usepackage{multirow}
\usepackage{setspace}
\usepackage{nomencl}
\usepackage{bookmark}
\usepackage{cases}

\makenomenclature

\usepackage{graphicx}

\modulolinenumbers[5]

\journal{Applied Mathematics Letters}

\begin{document}

\begin{frontmatter}

\title{An accurate lattice Boltzmann method for interface capturing}

\author[address]{Hong Liang}
\ead{lianghongstefanie@163.com}


\address[address]{Department of Physics, Hangzhou Dianzi University, Hangzhou, 310018, China}

\begin{abstract}
Accurately solving phase interface plays a great role in modeling immiscible multiphase flow system. In this letter, we propose an accurate interface-capturing lattice Boltzmann method from the perspective of the modified Allen-Cahn equation (ACE). The modified ACE is built based on the commonly used conservative formulation via the relation between the signed-distance function and the order parameter also maintaining the mass-conserved characteristic. A suitable forcing term is carefully incorporated into the lattice Boltzmann equation for correctly recovering the target equation. We then test the proposed method by simulating a typical interface-tracking problem of single vortex and demonstrate that the present model can be more numerically accurate than the existing lattice Boltzmann models for the conservative ACE, especially at a small interface-thickness scale.
\end{abstract}


\end{frontmatter}

\linenumbers

\section{Introduction}
Immiscible multiphase flows are universal in both nature and engineering applications. An indispensable event in modeling such flows is identifying and capturing the interface among multiple fluids. Phase field model~\cite{Shen, Kim1}, as one of diffuse interface methods, has received great success for interface capturing since it does not need to explicitly track the location of the interface and can handle large topological change naturally. In addition, the normal vector and curvature for the interface can be calculated directly from the phase field parameter. The well-known examples of the phase field models include the Cahn-Hilliard equation and the Allen-Cahn equation (ACE), both of which coupled with the hydrodynamic equations have been extensively used to describe complex interfacial dynamics~\cite{Shen, Kim1, Liang1} and liquid-solid phase transition~\cite{Sun}. However, from a mathematical point of view, the ACE only contains second-order derivatives compared with the Cahn-Hilliard equation including fourth-order spatial derivative and is numerically easier to be implemented. Within this context, several efforts
have been made to construct the efficient numerical approaches for the ACE within the finite-volume and finite-difference frameworks~\cite{Chiu, Mirjalili}.

The lattice Boltzmann (LB) method, as a mesoscopic numerical approach, has also been extensively employed to solve the ACE as the interface-capturing equation for simulating two-phase hydrodynamic phenomena. Geier {\it{et al.}}~\cite{Geier} proposed the first LB approach for the conservative ACE~\cite{Chiu} and found a higher numerical accuracy than the LB counterparts for the Cahn-Hilliard equation. However, some artificial terms appear in their recovered equation with the multiscale analysis procedure, and to remedy this drawback, some researchers~\cite{Wang, Ren, Liang2} have independently developed an improved LB model for the ACE by introducing a suitable source term on the time derivative of the order parameter. Also to recover the ACE correctly, Zu {\it{et al.}}~\cite{Zu} proposed another LB model for interface capturing, in which a nonlocal source term on the equilibrium distribution function was utilized. Recently, Zhang {\it{et al.}}~\cite{Zhang} elaborately designed a nondiagonal relaxation matrix and presented the multiple-relaxation-time LB models for recovering the correct ACE. It is worth noting that all aforementioned LB models were built based on the conservative ACE~\cite{Chiu}, which is superior in maintaining the mass-conservation property, but it could lead to the artificial distortion of the interface for long-time integrations due to the limited accuracy~\cite{Jain}. In this letter, we present an accurate LB method for interface capturing based on the improved conservative ACE~\cite{Jain}. The method is tested and compared with the existing LB models by simulating a benchmark problem of single vortex. Finally, a brief conclusion closes the paper.

\section{Accurate lattice Boltzmann method for interface capturing}
\subsection{Improved conservative Allen-Cahn equation}
Recently, the phase field model based on the ACE has been an intensively researched topic. The conventional ACE was originally developed for modeling solid-liquid phase change in crystalline growth~\cite{Allen} and the main issue of this original model lies in the lack of mass conservation. To enforce the mass conservation, Chiu and Lin~\cite{Chiu} reformulated the ACE by subtracting curvature-driven interface motion and then deduced the popular conservative ACE:
\begin{equation}
\frac{\partial\phi}{\partial{t}}+\nabla\cdot(\phi{\bf{u}})=\nabla\cdot{M}\left[\nabla\phi-\frac{4\phi(1-\phi)}{\epsilon}\frac{\nabla \phi}{|\nabla\phi|}\right],
\end{equation}
where $\phi$ is the order parameter taking 0 and 1 in the bulk phases and varying smoothly across the interface, $M$ is the mobility, ${\epsilon}$ is the interface thickness scale parameter. This ACE is able to strictly preserve the global mass conservation and has been successfully employed by several traditional numerical solvers~\cite{Chiu, Mirjalili} and also within the LB framework~\cite{Geier, Wang, Ren, Liang2, Zu, Zhang}. Nonetheless, it sometimes would suffer from two drawbacks, one of which is the artificial distortion of the interface for long-time integrations and the second is the degraded accuracy at enough small interface thickness~\cite{Jain}. To overcome these limitations, Jain~\cite{Jain} recently derived an improved conservative ACE by using the relation between the signed-distance function originating from the level-set method and the order parameter. We first present the analytical distribution of the order parameter at the equilibrium state as
\begin{equation}
\phi=\frac{1}{2}+\frac{1}{2}\tanh\left(\frac{2s}{\epsilon}\right),
\end{equation}
where $s$ represents the coordinate normal to the interface, and then the normal derivative of the order parameter can be expressed by
\begin{equation}
|\nabla\phi|=\frac{d\phi}{ds}=\frac{4\phi(1-\phi)}{\epsilon}.
\end{equation}
Let $\psi$ denote the signed-distance function from the interface such that having the relation $\psi=s(\phi)-s(\phi=0.5)$, the derivative of the order parameter with respect to $\psi$ can be written as
\begin{equation}
\frac{d\phi}{d\psi}=\frac{4\phi(1-\phi)}{\epsilon}.
\end{equation}
Integrating the above equation, one can eventually deduce the relation between the order parameter and the signed-distance function as
\begin{equation}
\phi=\frac{e^{(\psi/\epsilon)}}{1+e^{(\psi/\epsilon)}}=\frac{1}{2}+\frac{1}{2}\tanh\left(\frac{2\psi}{\epsilon}\right)=,~~or~~\psi=\epsilon \ln\left(\frac{\phi}{1-\phi}\right),
\end{equation}
which further leads to
\begin{equation}
\frac{4\phi(1-\phi)}{\epsilon}=\frac{1}{\epsilon}\left[1-\tanh^2\left(\frac{2\psi}{\epsilon}\right)\right]
\end{equation}
and the unit normal vector computed by
\begin{equation}
\mathbf{n}=\frac{\nabla \phi}{|\nabla \phi|}=\frac{\nabla\psi}{|\nabla\psi|}.
\end{equation}
Join~\cite{Jain} suggested replacing the sharpening flux term $\phi(1-\phi)$ and the unit normal vector $\frac{\nabla \phi}{|\nabla \phi|}$ in Eq. (1) with the relations (6) and (7), since $\psi$ is a better-behaved function in comparison to $\phi$ avoiding any jumps or discontinuities. With these
modifications, the improved conservative ACE can be formulated by
\begin{equation}
\frac{\partial\phi}{\partial{t}}+\nabla\cdot(\phi{\bf{u}})=\nabla\cdot{M}\left\{\nabla\phi-\frac{1}{\epsilon}\left[1-\tanh^2\left(\frac{2\psi}{\epsilon}\right)\right]\frac{\nabla \psi}{|\nabla\psi|}\right\},
\end{equation}
All the terms in Eq. (8) are in divergence form such that the improve Allen-Cahn phase field model is able to preserve the mass
conservation property. In addition, the modified ACE can be still deemed as a single-scalar convection-diffusion equation since the signed-distance function $\psi$ in Eq. (8) can be determined by the the algebraic relation (5), in this case the model is also cost-effective and easy to be implemented.

\subsection{Lattice Boltzmann method}
All existing LB methods~\cite{Geier, Wang, Ren, Liang2, Zu, Zhang} were constructed stating from the conservative ACE as Eq. (1) and these methods would surely inherit these weaknesses of the phase field model described above. We attempt to present the first LB approach for solving the improved conservative ACE. Applying the BGK collision operator, the LB evolutional equation with a forcing distribution function can be written as~\cite{Chai}
\begin{equation}
{f_i}({\bf{x}} + {{\bf{c}}_i}\delta_t,t + \delta_t) -
{f_i}({\bf{x}},t) =  -\frac{1}{\tau}\left[ {{f_i}({\bf{x}},t) -
f_i^{eq}({\bf{x}},t)} \right] + {\delta_t}(1-\frac{1}{2\tau}){F_i}({\bf{x}},t).
\end{equation}
where ${f_i}({\bf{x}},t)$ is the particle distribution function, $f_i^{eq}({\bf{x}},t)$ is the equilibrium distribution function, $\tau$ is the dimensionless relaxation factor, $\delta_t$ is the time step, ${F_i}({\bf{x}},t)$ is the forcing distribution function. The equilibrium distribution function must be chosen to satisfy the moment conditions required for recovering the correct macroscopic behaviour. To match the target ACE, we use the following equilibrium distribution function,
\begin{equation}
f_{i}^{eq}={\omega_i}\phi\left(1+\frac{{\bf{c}}_i\cdot\mathbf{u}}{c_s^2}\right),
\end{equation}
where ${\omega_i}$ is the weighting coefficient, ${c_s^2}$ is the sound speed and ${\bf{c}}_i$ is the discrete velocity determined by the discrete-velocity lattice model. Here we utilize the D2Q9 lattice structure for two-dimensional space, where $\omega_i$ is given by $\omega_0=4/9$, $\omega_{1-4}=1/9$, $\omega_{5-8}=1/36$, $c_s^2=1/3$, and the discrete velocity $\textbf{c}_i$ can be referred to Ref.~\cite{Liang2}. The diffusion term on the $\psi$ in Eq. (8) is regarded as the source term and should be carefully incorporated into the LB algorithm owing to the possibly produced discrete lattice effect. We take into account the discrete lattice effect and construct an efficient forcing distribution function,
 \begin{equation}
{{F}_i} ={\omega_i}{{{{\bf{c}}_i} \cdot
\left[{\partial _t}(\phi{\bf{u}})+c_s^2\frac{\nabla \psi}{|\nabla\psi|}\frac{1-\tanh^2(\frac{2\psi}{\epsilon})}{\epsilon}\right]} \over {c_s^2}},
\end{equation}
where the time derivative is included to remove the unwanted term similar to the strategy in the LB modeling for the Cahn-Hilliard equation~\cite{Liang1}. The order parameter in the LB method is calculated by
\begin{equation}
\phi = \sum\limits_i {{f_i}}.
\end{equation}
Following the Chapman-Enskog procedure as Ref.~\cite{Wang} , we can demonstrate the present LB method can recover the improved conservative ACE exactly with the mobility given by ${M} =c_s^2(\tau-0.5)\delta_t$. In practice, the LB method contains some derivatives needed to be numerically evaluated. We use the explicit Euler scheme~\cite{Wang, Liang2} to compute the time derivative and considering the consistency in comparison with the available LB models~\cite{Geier, Zu}, the commonly used second-order isotropic discretization~\cite{Liang1} is applied to compute the gradient operator. Another crucial event in implementing the LB method requires the careful computations of the signed-distance function $\psi$ and the unit normal vector $\mathbf{n}$, as they would approach infinities in bulk phase region. To avoid this problem, an extremely small value $\varepsilon$ should be added by
\begin{equation}
\psi=\epsilon \ln(\frac{\phi+\varepsilon}{1-\phi+\varepsilon}),~~~\mathbf{n}=\frac{\nabla\psi}{|\nabla\psi|+\varepsilon},
\end{equation}
where $\varepsilon=e^{-20}$ is used in the numerical experiment. Additionally, the order parameter $\phi$ could not be always remaining bounded between 0 and 1 due to the slight diffusion effect, then evaluating $\psi$ by relation (13) could produce physically unrealistic values. To solve this problem, one can clip $\phi$ between 0 and 1 before computing $\psi$ by replacing $\phi$ in Eq. (13) with $\phi_{clip}=\min(\max(0,\phi),1))$. This will not create any conservation issues because $\phi$ is not directly updated, it is only clipped while computing $\psi$.

\section{Numerical test}

\begin{figure}
\centering
\includegraphics[width=4.0in,height=2.8in]{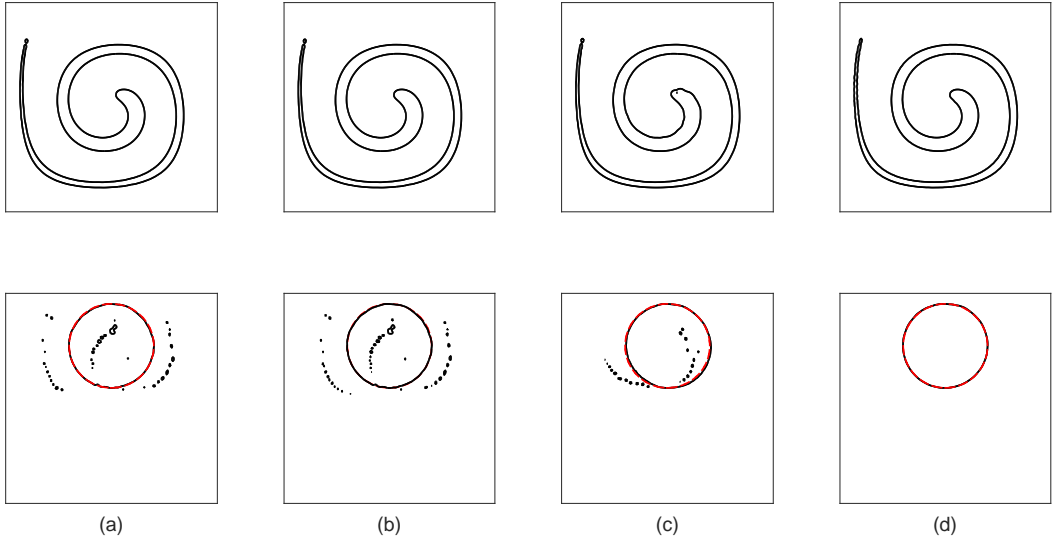}~~~~~~~~~~~~~~~~~~~~~~~~~~~~~~~~~~~~~~~~~
 \tiny\caption{The numerical predictions of disk's shape with different LB models, (a) the model of Geier~\cite{Geier}, (b) the model of Wang~\cite{Wang}, (c) the model of Zu~\cite{Zu}, (d) the present LB method. The red dotted curve represents the analytical profile of the disk interface at one period.}
\end{figure}

\begin{figure}
\centering
\includegraphics[width=3.0in,height=2.4in]{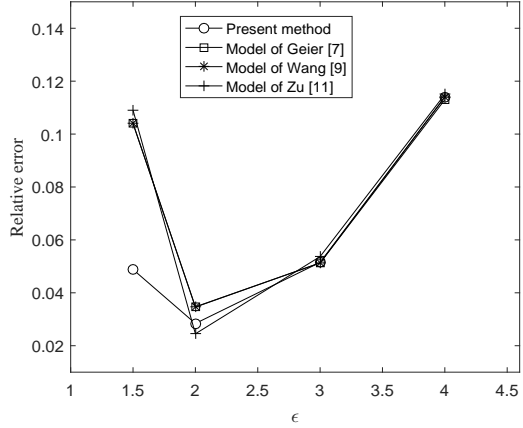}
\tiny\caption{The dependence of the interface-capturing relative error on the interfacial thickness with different LB models.}
\end{figure}

A typical numerical test of time-reversed single vortex~\cite{Rider} is carried out to examine the accuracy and efficiency of the present LB scheme for interface capturing with a specific velocity distribution. To quantitatively describe the accuracy of the present method and compare with the existing LB models~\cite{Geier,Wang, Zu}, the $L_1$-norm relative error of the order parameter between numerical and analytical solutions is used. Initially, a circular disk is placed a periodic domain evenly divided by $L\times L$ lattice cells with $L=200$ and its radius occupies a size $R=0.2L$ centered at $(x_c,~y_c)=(0.5L,~0.75L)$. To be smoothly change across the interface, the initial profile of the order parameter can be given following the hyperbolic tangent function: $\phi(x,y)=0.5+0.5\tanh 2[{R-\sqrt{(x-x_c)^2+(y-y_c)^2}}]/{\epsilon}$. The disk would be deformed and stretched by imposing a nonlinear velocity field $(u,v)$,
\begin{equation}
u=U_0\sin^2\frac{\pi x}{L}\sin\frac{2\pi y}{L}\cos\frac{\pi t}{T},~~
v=-U_0\sin^2\frac{\pi y}{L}\sin\frac{2\pi x}{L}\cos\frac{\pi t}{T},
\end{equation}
where $U_0$ is the reference velocity, $t$ is the evolutional time scaled by $L/U_0$, $T$ is the periodic time. The time-reversed function added in Eq. (14) is to change the sign of the velocity field at $T/2$ such that the disk would return to its initial configuration at a periodic time $T$. This test poses some challenges since the interface has a large deformation with topological change, especially at an increasing time $T$. Note that, most of previous numerical test was limited to $T=4$~\cite{Chiu, Jain} and we consider a longer periodic time $T=6$. The remaining physical parameters are given as $U_0=0.04$ and $M=0.001$. We first simulated this case with a mild interface thickness $\epsilon=3$ and find that all these LB models are capable of achieving the comparative numerical accuracy in solving phase interface. We further consider the example with a small interface thickness $\epsilon=1.5$ since the phase interface is physically sharp in the continuum limit. Figure 1 displays the numerically predicted disk's patterns during one period with different LB models. It can be observed that all existing LB models~\cite{Geier, Wang, Zu} produce unphysically ruptured small droplets at the end of the filament for half period and also quantities of undesirable distortions at the vicinity of the disk for one period, while the present LB approach is able to derive a more stable and accurate interface. The predicted shape of the disk at one period exactly lines up over the analytical profile. We also performed several simulations with a series of interface thicknesses and calculated the global relative errors in Fig. 2. It is found that the present method generally derives the smallest numerical error at a small interface thickness and would provide an improved accuracy of interface capturing within the LB framework for the ACE.

\section{Conclusion}
In this letter, a fresh LB method is developed for interface capturing based on an alternative conservative ACE. This novel formulation can be viewed as a modified version of the popular conservative ACE by using the relation between the signed-distance function and the order parameter. Since the signed-distance function is a better-behaved quantity across the interface, the modified ACE achieves a better accuracy in interface capturing and also maintains the mass-conserved property. In order to recover the reformulated interfacial equation, a proper discrete forcing term is exactly designed and incorporated into the present LB framework. We also validated the present LB approach by simulating a stringent benchmark problem of single vortex and further compared it with several existing LB models for conventional ACE. It can be found that the present LB method enables to significantly improve numerical accuracy and stability of interface capturing at a reducing interface thickness.

\section{Acknowledgements}

This work is financially supported by the National Natural Science Foundation of China (Grant No. 11972142).

\end{document}